\documentclass[amsmath,amssymb,aps,prl,reprint,floatfix,superscriptaddress]{revtex4-2}
\usepackage{xcolor}
\usepackage{graphicx}
\usepackage{bm}
\usepackage{booktabs} 
\usepackage{appendix}
\usepackage{xr-hyper}
\usepackage{hyperref}
\hypersetup{colorlinks=true,linkcolor=blue,citecolor=blue,urlcolor=blue}
\begin{document}
\title{Heat transport in weakly anharmonic Fermi-Pasta-Ulam-Tsingou chains}
\author{Kiratholly Nandakumar Madhav Sharma}
\affiliation{Univ. Grenoble Alpes, CNRS, LPMMC, 38000 Grenoble, France}
\author{Stefano Mossa}
\affiliation{Univ. Grenoble Alpes, CEA, IRIG-MEM-LSim, 38054 Grenoble, France}
\author{Jean-Louis Barrat}
\affiliation{Univ. Grenoble Alpes, CNRS, LIPhy, 38000 Grenoble, France}
\author{Markus Holzmann}
\affiliation{Univ. Grenoble Alpes, CNRS, LPMMC, 38000 Grenoble, France}
\date{\today}
\begin{abstract}
We investigate anomalous heat transport in the one-dimensional $\beta$-Fermi--Pasta--Ulam--Tsingou chain in the weakly anharmonic regime using large-scale equilibrium molecular dynamics. By resolving the heat current into phonon-mode contributions, we show that its long-time autocorrelation is controlled by diagonal mode correlations, whose decay is quantitatively determined by the corresponding phonon damping rates. Extrapolation to the thermodynamic limit gives $\Gamma_k \propto (\beta T)^{4/3} k^{5/3}$, with a subleading $k^2$ correction. The latter produces an extended pre-asymptotic regime, $C(t) \sim t^{-4/5}$, before the asymptotic decay $C(t) \sim t^{-3/5}$ is reached, leading to $\kappa(N) \sim N^{2/5} + O(N^{1/5})$. At weak anharmonicity, off-diagonal mode correlations persist up to a characteristic time $\tau_{\mathrm{od}} \sim (\beta T)^{-2}$, while increasingly large systems are required to resolve the long-wavelength phonon contribution. Weak anharmonicity therefore does not introduce a distinct asymptotic transport mechanism, but shifts the onset of the phonon-dominated regime to progressively longer times and larger length scales.
\end{abstract}
%
%
\maketitle
\paragraph{Introduction.} In one-dimensional momentum-conserving lattice models, thermal conductivity diverges with system size, $N$, as $\kappa(N) \sim N^{\nu}$,
%
%
with $0<\nu<1$ indicating anomalous heat transport~\cite{lepri1998anomalous,lepri2026anomalous, 10.1063/1.1854281, wang2011power, PhysRevLett.78.1896, takatsu2024large}. The $\beta$-Fermi--Pasta--Ulam--Tsingou ($\beta$-FPUT) chain is a paradigmatic model for investigating this behavior. 
Its dynamics are governed by the Hamiltonian,
\begin{equation}
{\cal H} = \sum_{j=1}^{N} \left[\frac{v_j^2}{2}
+ \frac12 \left(x_{j+1}-x_j\right)^2 
+ \frac{\beta}{4} \left(x_{j+1}-x_j\right)^4
\right],
\label{eq:H_betaFPUT}
\end{equation}
where $x_j$ and $v_j$ denote the displacement and velocity of particle $j$, respectively, with appropriate boundary conditions imposed. At temperature $T$, the effective strength of the anharmonicity is controlled by $\beta T$.
Whereas hydrodynamic fluctuation~\cite{PhysRevE.90.012124}, self-consistent mode-coupling  theory~\cite{Delfini_2007}, and renormalization group arguments~\cite{Narayan2002} suggest $\nu=1/2$, predictions based on kinetic theory~\cite{Lepri1998,Lukkarinen2008,Pereverzev2003,Dematteis2020}, typically formulated in the weakly anharmonic regime, provide $\nu=2/5$. 

Since the validation of the different pictures necessarily relies on numerical simulations, finite-size and finite-time effects represent a major obstacle, especially when the focus is exclusively on asymptotic behavior.
Early non-equilibrium molecular dynamics simulations by Lepri \textit{et al.}~\cite{PhysRevLett.78.1896} suggested $\nu=1/2$. Subsequent equilibrium studies~\cite{lepri1998anomalous,10.1063/1.1854281}, however, reported a smaller value, close to $0.37$. More recent large-scale equilibrium and non-equilibrium simulations~\cite{wang2011power,takatsu2024large} with $\beta T \approx 1$ revealed a nontrivial, non-monotonous crossover in the apparent scaling of $\kappa(N)$ with  $\nu\simeq 2/5$ for the largest system sizes, reaching $N \sim 10^7$~\cite{takatsu2024large}. Although $\nu =2/5$ seems to support the validity of kinetic theory, based on Boltzmann equation, direct numerical studies of $\nu$ in the weakly anharmonic regime, $\beta T \lesssim 0.1$, are less conclusive~\cite{10.1063/1.1854281,Gill-Comeau2014}, questioning the universality of the power law exponent~\cite{Lepri2020}.

Here, we perform large-scale equilibrium molecular dynamics calculations focusing on the weakly anharmonic limit. Analyzing the different contributions to the heat current auto-correlation function, we show that the divergence of the thermal conductivity can be quantitatively related to the phonon frequency, $\omega_k$, and damping, $\Gamma_k$, via the Peierls-Boltzmann equation. We highlight the different time and length scales needed to reach the asymptotic region, and discuss the sub-leading corrections leading to transient behavior in the correlation functions. As a result, a smoothly varying  effective exponent appears, approaching non-monotonously the asymptotic value. Our results show that weakly and strongly anharmonic region join smoothly, both yielding $\nu \simeq 2/5$. 
%
%
\paragraph{Methods and results.} We have performed constant-energy equilibrium molecular dynamics (EMD) simulations of $N$ particles in a periodic chain using the Velocity-Verlet algorithm, starting from initial configurations close to thermal equilibrium. 
Weakly anharmonic systems are usually described in terms of phonons. We have calculated phonon frequencies, $\omega_k$, and damping rates, $\Gamma_k$, by analyzing the normal mode correlation functions, $c_k(t)=\langle q_k^*(t)q_k(0)\rangle$, where $q_k$ is given by the Fourier transform  of the particle displacements, with $k \in [-\pi,\pi]$. 
In the harmonic limit, the normal modes oscillate undamped with frequencies $\omega_k^0=|2 \sin(k/2)|$. 
For $\beta T>0$, we obtain $\omega_k$ and $\Gamma_k$ from a Lorentzian fit of the peak in the spectral function of the normal modes.
\begin{figure}[t]
\centering
\includegraphics[width=1.0\columnwidth]{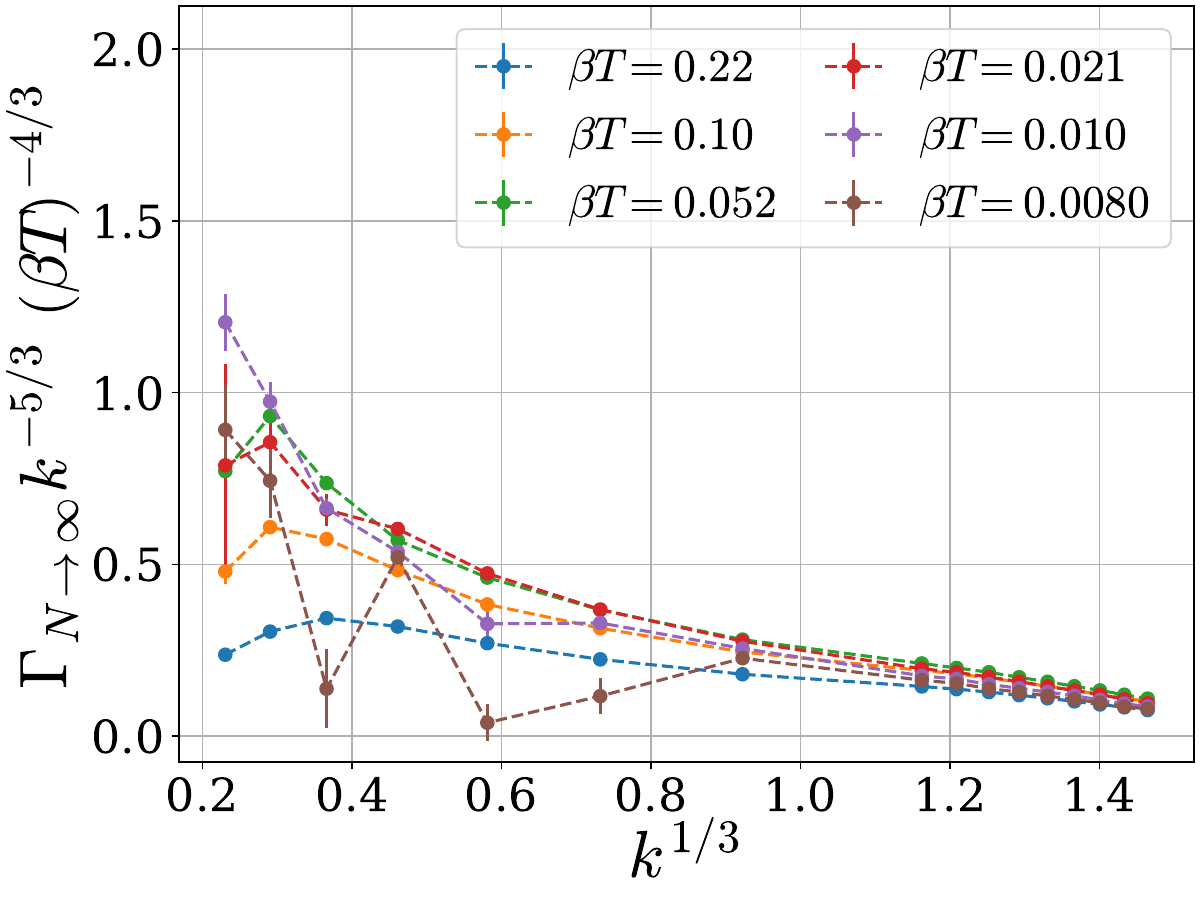}
\caption{Phonon damping rates extrapolated to the thermodynamic limit $ N \rightarrow \infty $, normalized by the leading $k^{5/3}$ and $(\beta T)^{4/3}$ scaling (see Eq.~(\ref{eq:Gamma_k1})). }
\label{fig:Gamma_k}
\end{figure}

The frequencies are to leading order renormalized to $\omega_k=\sqrt{\eta} \omega_k^0$, where $\eta \simeq 1/2 + \sqrt{1/4 + 3 \beta T}$ is an overall renormalization factor (see the supplementary material). Whereas finite-size effects are not relevant for the leading-order frequency renormalization, phonon damping rates require careful extrapolation to the thermodynamic limit. For any system size, $N$, the typical time needed for phonons of wave vector $k$ to traverse the entire chain is $t_N \sim N/|v_k|$, with $v_{k}=d \omega_k/dk$ the group velocity. Phonons with  $\Gamma_k t_N \gtrsim 1$ therefore relax almost completely before boundary effects could significantly interfere. This condition can be satisfied for any finite value of $k$ by considering sufficiently large system sizes such that $N \gtrsim |v_k| / \Gamma_k$, a condition which can be checked self-consistently. Remaining finite-size effects are hence extrapolated based on a polynomial in $1/N$. 

Our main results for the damping rate $\Gamma_k$, extrapolated to the thermodynamic limit, are shown in Fig.~\ref{fig:Gamma_k}. 
We extract an approximate scaling law,
%
\begin{equation}
\Gamma_k \approx
\Gamma_0 (\beta T)^{4/3}\frac{ k^{5/3}}{1 + (k/k_0)^{1/3}},
\label{eq:Gamma_k1}
\end{equation}
which characterizes reasonably well our data in the range $8 \times 10^{-3} \le  \beta T \le 0.22$, with small systematic deviations visible at the largest anharmonicity strengths, $\beta T = 0.10$ and $0.22$. Note that while the $k^{5/3}$ dependence is characteristic of damping determined by using the Fermi's golden rule in second-order perturbation theory~\cite{Pereverzev2003, Lukkarinen2008, Dematteis2020}, the overall strength $(\beta T)^{4/3}$ cannot be explained within perturbation theory. Such a leading-order behavior in $k$ and $\beta T$ for the damping rate has been already suggested based on mode coupling theory in~\cite{Lepri1998}, although the derivation has been later criticized as inconsistent in~\cite{Narayan2002,Delfini_2007}. 

However, for any finite $N$, the lowest modes with $k$ below a threshold $k_m$ will travel essentially ballistically whenever $2\Gamma_k N/v_k \lesssim 1$.
From our results above, setting $2\Gamma_{k_m} N/v_{k_m} = 1$, we find $ k \lesssim k_m \sim (\beta T)^{-4/5} N^{-3/5}$. In the weakly anharmonic limit, large system sizes, $N \gg (\beta T)^{-4/3}$, are therefore required to reduce the fraction of ballistic modes and to efficiently approach the thermodynamic limit behavior.
\begin{figure}[b]
\centering
\includegraphics[width=1.0\columnwidth]{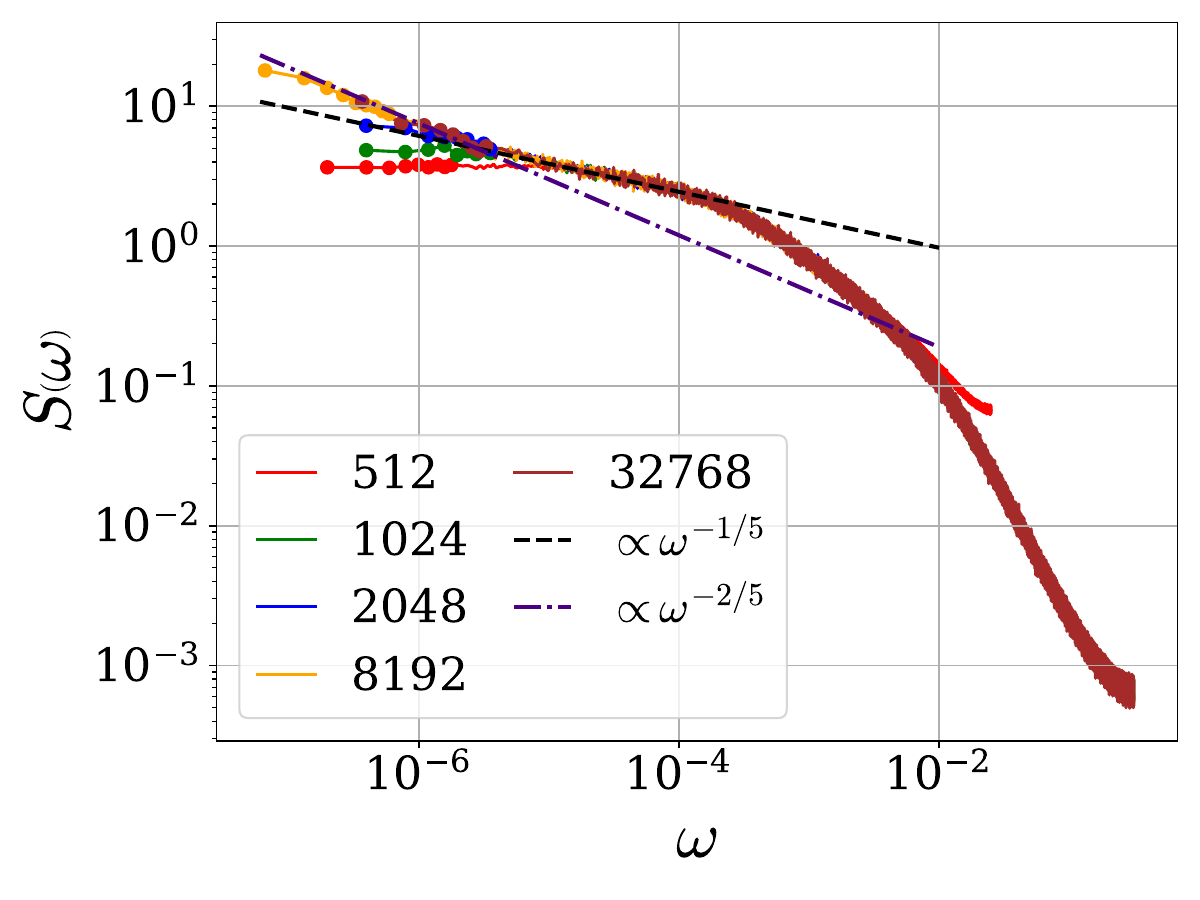}
\caption{
Power spectrum $S(\omega)$ at the indicated values of $N$ for $\beta T=10^{-1}$. The dashed lines are guides to the eye proportional to $\omega^{-1/5}$ and $\omega^{-2/5}$.}
\label{fig:SofOmega}
\end{figure}

In thermodynamic equilibrium, the thermal conductivity can be obtained from the current-current autocorrelation function within the Green-Kubo linear response theory as,
\begin{equation}
\kappa =
\frac{1}{ k_{\mathrm{B}} T^{2}}
\lim_{t \to \infty} \lim_{N\to \infty} \frac{1}{N}\int_{0}^{t}
dt' \big\langle J(t')J(0)\big\rangle. 
\label{eq:GK_relation_for_kappa}
\end{equation}
%
%
In our simulations, we set $k_B = 1$ and calculate the energy current as, 
\begin{equation}
J(t) = \frac{1}{2} \sum_{i=1}^{N} \left[ v_{i+1}(t) + v_i(t) \right] 
\left[x_{i + 1} - x_{i} + \beta \left(x_{i + 1} - x_{i}\right)^3 
\right].
\label{eq:Heatflux_formula}
\end{equation} 
%
%
Based on the analytical calculations of~\cite{Lepri1998, Narayan2002,Pereverzev2003,Delfini_2007,Lukkarinen2008,Dematteis2020,PhysRevE.90.012124}, $C(t)=\langle J(t)J(0)\rangle/(NT^2)$ is expected to decay with a power law slower than $t^{-1}$, so that $\kappa$ diverges in the thermodynamic limit. To match the divergence $\kappa(N) \sim N^\nu$ observed in non-equilibrium calculations for finite systems, one usually replaces the upper limit $t$ in the integral of Eq.~\eqref{eq:GK_relation_for_kappa} with a system-size-dependent cutoff, $t_N=\lim_{k\to 0} N/|v_{k}| \simeq N$ \cite{wang2011power,hou_phonon_2012, Pereverzev2003, lepri1998anomalous}. From this argument, the auto-correlation function of the total current should asymptotically decay as $t^{-(1-\nu)}$ in order to be compatible with the expected $\kappa(N) \sim N^\nu$.

Extrapolations of the current correlation function to the thermodynamic limit are best obtained in terms of the corresponding power spectrum,
\begin{equation}
S(\omega) = 2 \int_{0}^{\infty} dt\, C(t)  \cos \left(\omega t\right), 
\end{equation}
which, for $\omega \to 0$, is therefore expected to behave as $\omega^{-\nu}$. We show $S(\omega)$ in Fig.~\ref{fig:SofOmega}, for $\beta T=10^{-1}$ and system sizes between $N=512$ and $32768$. We see that the larger system sizes collapse reasonably well on the same curve, indicating that the thermodynamic limit has been reached within the shown frequency window. A power law behavior $\sim \omega^{-2/5}$ emerges for $\omega \lesssim 10^{-6}$.
\begin{figure}[b]
\centering
\includegraphics[width=1.0\columnwidth]{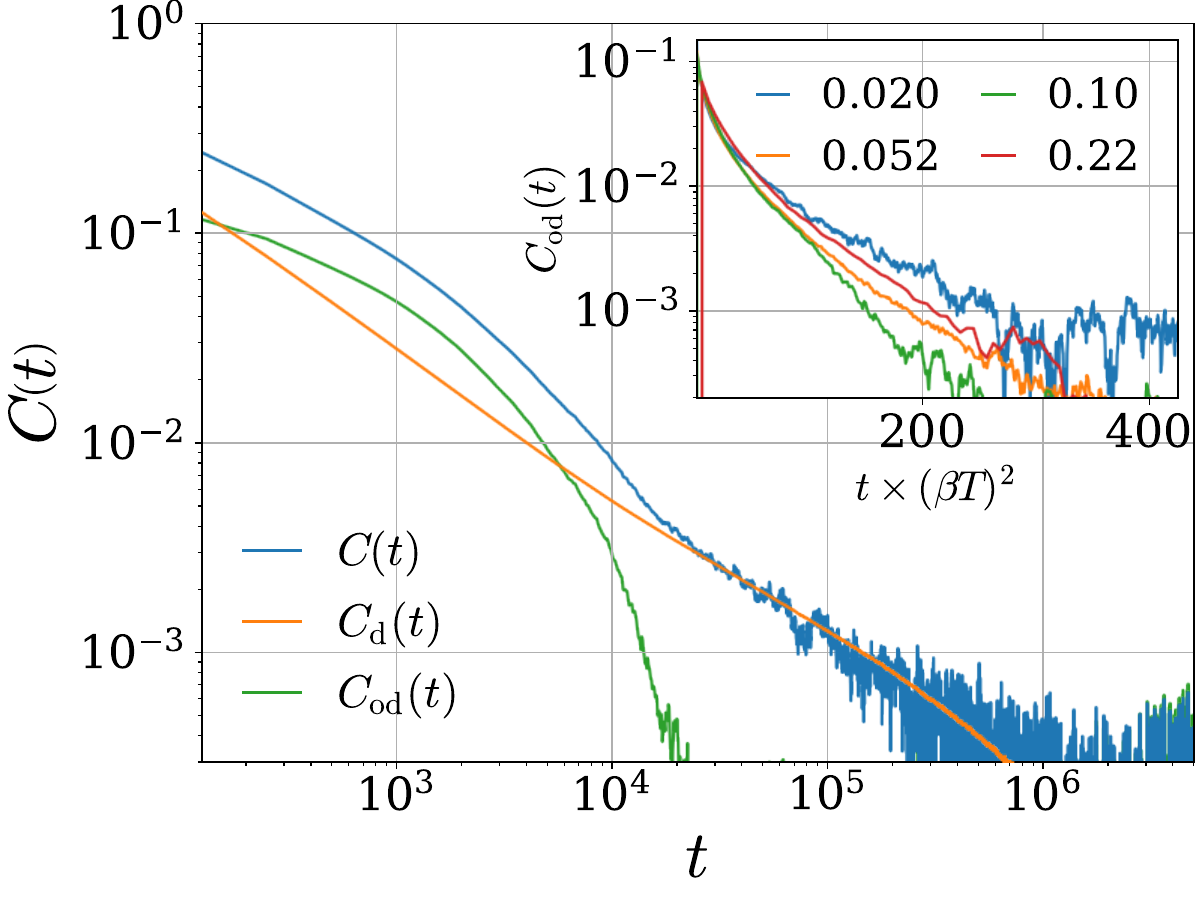}
\caption{Diagonal and off-diagonal contributions to the harmonic heat-current autocorrelation function for $N=8192$ and $\beta T=10^{-1}$. Beyond the crossover time $t_c=6\times10^3$, the diagonal contribution dominates, marking the onset of the asymptotic power-law regime. In the inset we show $C_{\mathrm{od}}(t)$, for $N=8192$ and the indicated values of $\beta T$, with time rescaled as $t(\beta T)^2$. 
}
\label{fig:Harmonic_Analysis}
\end{figure}

To understand in depth this behavior, we focus on the  correlation  function of the {\em harmonic} currents, $C^h(t)=\langle \eta J^h(t) \eta J^h(0)\rangle/(NT^2)$, where $J^h(t)$ is the harmonic part of Eq.~(\ref{eq:Heatflux_formula}), simply obtained by setting $\beta=0$. Note that we have inserted a factor $\eta^2$, so that $C^h(t)$ matches $C(t)$ to leading order in $\beta T \to 0 $~\cite{lepri1998anomalous}. We now decompose the harmonic current into phonon modes as $\eta J^h =2\sum_{k>0}^\pi J_k$, with
\begin{equation}
J_k(t) = \frac{i v_k \omega_k}{2}
\left[
\dot{q}_k(t)q_{k}^*(t)
-
\dot{q}_{k}^*(t)q_k(t)
\right],
\end{equation}
where $v_k=d\omega_k/dk$ is the group velocity corresponding to the re-normalized frequency $\omega_k=\sqrt{\eta} \omega_k^0$. Then, we separate the diagonal and off-diagonal contributions~\cite{Gill-Comeau2014} writing $C^h(t)=C_{d}(t)+C_{od}(t)$ with,
\begin{equation}
C_{d}(t) = \frac{4}{NT^2}\sum_{k > 0}^{\pi} \left\langle J_k(t)J_k(0)\right\rangle.
\label{eq:off_diagonal_contribution}
\end{equation}
From Fig.~\ref{fig:Harmonic_Analysis} we see that, for $\beta T=10^{-1}$ and $N=8192$, the off-diagonal contribution dominates over the diagonal one over a significantly long time, until it decays approximately exponentially. Simultaneously, the diagonal term remains close to a power law up to large times, as similarly observed in~\cite{Gill-Comeau2014}.

In the inset, we show that the off-diagonal contributions decay exponentially with a typical time scale $\tau_{od} \sim (\beta T)^{-2}$, as expected from perturbation theory  in the small anharmonicity limit. Therefore, in the long-time limit, $t \gg (\beta T)^{-2}$, the current correlation function is entirely described by the diagonal term, which decays as a power law in the limit of large system sizes. Since we may expect important finite-size errors at long times $t \gtrsim t_N \sim N$, one might have to increase system sizes $N \gg (\beta T)^{-2}$ in the weak anharmonic limit. On the other hand, decreasing $\beta T$ for any fixed value of $N$, the time correlation will be dominated by the off-diagonal term, as reported in~\cite{Gill-Comeau2014}.

\begin{figure}[t]
\centering
\includegraphics[width=1.0\columnwidth]{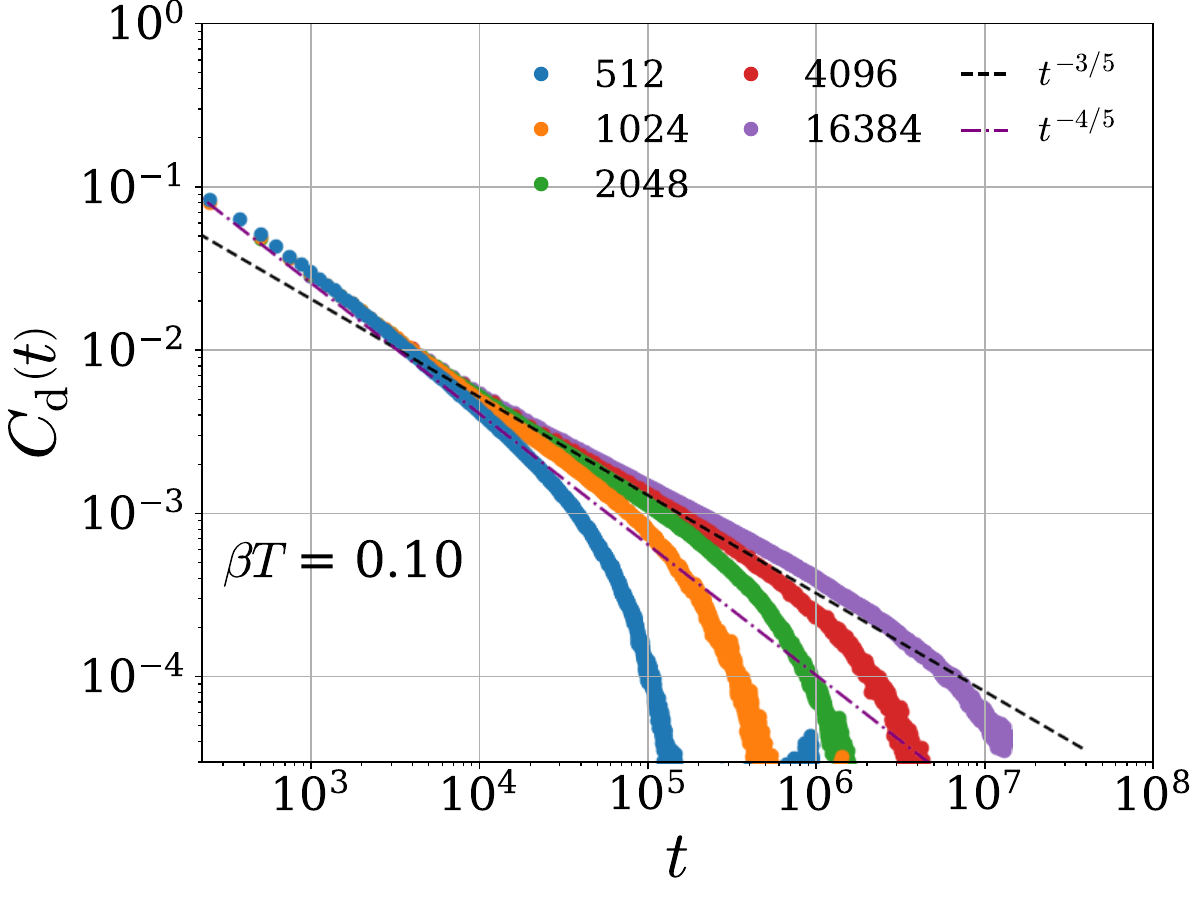}
\caption{Diagonal contribution to the heat-current autocorrelation function for $\beta T=10^{-1}$ and the indicated values of $N$. 
The dashed lines indicate the reference power laws $t^{-4/5}$ and $t^{-3/5}$.}
\label{fig:Cdoftau}
\end{figure}

In Fig.~\ref{fig:Cdoftau} we examine in detail the diagonal contribution $C_d(t)$ for different system sizes. The correlations first decay approximately algebraically, before turning into exponential at some larger time,  $t_m$. Since $t_m$ increases with  system size, the algebraic decay will dominate the large-time behavior in the thermodynamic limit. Still, extracting the exponent of the power law needs some care, as the effective exponent changes slowly from an apparent $t^{-4/5}$ behavior to $t^{-3/5}$.

\begin{figure}[t]
\centering
\includegraphics[width=\columnwidth, height=0.85\columnwidth]{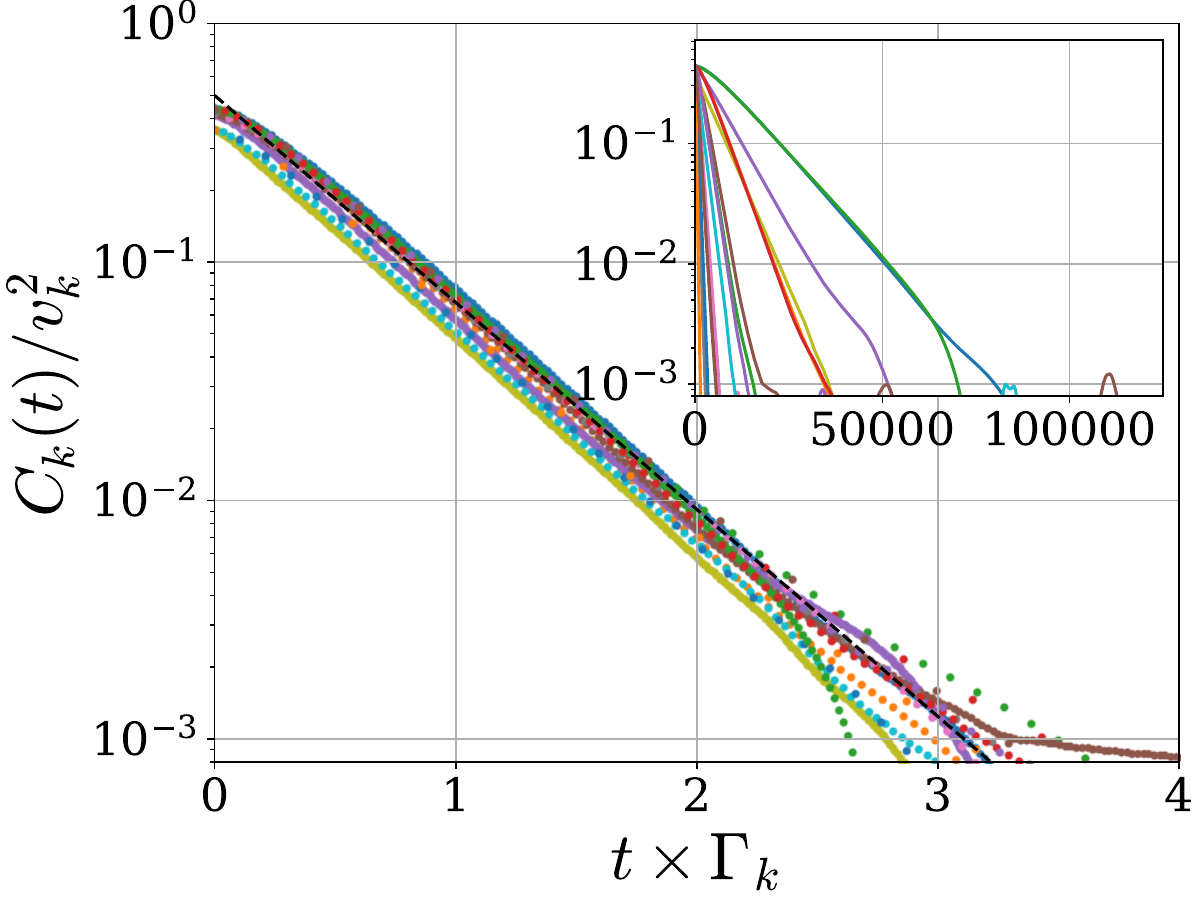}
\caption{ Mode-current autocorrelation functions for representative phonon modes spanning system sizes $N=8192$--$32768$, anharmonicity strengths $\beta T=0.052$--$0.22$, and wave vectors $k=0.024$--$0.20$. Rescaling time by the corresponding phonon damping rate $\Gamma_k$ produces a collapse of the data onto a common curve. 
The black dashed line represents the reference exponential $0.5 \, e^{-2 x}$. 
In the inset, we show the unscaled data.}
\label{fig:Mode_Current_Auto_Correlation_function}
\end{figure}

To obtain more insight, we have determined the decay of the individual modes $k$, $C_k(t)=\langle J_k(t)J_k(0) \rangle/T^2 \simeq \eta \,v_k^2 e^{-\gamma_k t}/2$ at large times. 
In Fig.~\ref{fig:Mode_Current_Auto_Correlation_function} we show $C_k(t)$ for different modes, non-linearities, and system sizes.  Within our numerical uncertainties, the decay is determined by the phonon lifetime, $\gamma_k = 2 \Gamma_k$.
Indeed, as shown in Fig.~\ref{fig:Mode_Current_Auto_Correlation_function}, all curves rescaled by $v_k^2$ essentially collapse to a single exponential as a function of $t \Gamma_k$, where $\Gamma_k$ is the corresponding phonon damping rate. This suggests a simple expression for the diagonal part of the correlation function,
\begin{equation}
C_d(t) \simeq \frac{2 \eta }{N} \sum_{k>0}^\pi v_k^2 e^{-2 \Gamma_k t},
\label{eq:c_diagonal}
\end{equation}
which directly connects the long-time behavior of the current-current correlation function to equilibrium phonon frequencies and damping rates.

We can now analyze and clarify the detailed behavior of $C_d(t)$ employing the phonon decay rates of Eq.~(\ref{eq:Gamma_k1}). In particular, we can perform the extrapolation to the thermodynamic limit, $C_d^\infty(t) =\lim\limits_{N \to \infty}C_d(t)$, and obtain, 
\begin{eqnarray}
C_d^\infty(t)& = &  \eta \int_0^\pi \frac{dk}{\pi} v_k^2 e^{-2 \Gamma_k t} \\
&\approx &   \eta^2 \int_0^\pi \frac{dk}{\pi} \cos^2(k/2) e^{-2\Gamma_0 (\beta T)^{4/3}
(k^{5/3}-k_0^{-1/3} k^2)t}.
\nonumber
\end{eqnarray}
Asymptotically, for $t \gg t_0  = [2\Gamma_0]^{-1} (\beta T)^{-4/3}$, we find,
\begin{equation}
C_d^\infty(t \gg t_0)
\simeq \frac{3\eta^2}{5 \pi} \Gamma\left(\frac{3}{5}\right)(t/t_0)^{-3/5}
 + \mathcal{O}\left((t/t_0)^{-4/5} \right)
\label{eq:Heat_current_autocorrelation_in_thermodynamic_limit}
\end{equation}
%
where $\Gamma(3/5)\simeq 1.489$, for detailed derivation see supplementary material.
%
For any finite $N$, at very large times, finite-size effects will appear, so $C_d(t)$ is eventually dominated by a few modes and decays exponentially due to mode discretization.
This will roughly occur for $t \gtrsim t_{max}$ when $C_d^\infty(t_{max}) \sim \eta^2/N$, or $t_{max} \sim  (N/2\pi)^{5/3} t_0$. Conversely,  for large enough systems where $t_{max} \gg t_N \sim N$, the diagonal correlations
can be quantitatively approximated by their thermodynamic limit, $C_d(t) \simeq C_d^\infty(t)$, up to very large times $t \lesssim t_{max} \sim N^{5/3}$.

Let us now reconsider the thermal conductivity,
\begin{equation}
\kappa(N) = \kappa_{d}(N) + \kappa_{od}(N),
\end{equation}  
discussing separately the diagonal and off-diagonal contributions,
\begin{equation}
\kappa_{d}(N)= \int_0^{t_N} dt'\, C_{d}(t').
\end{equation}
and, analogously, $\kappa_{od}(N)$. 
%
%
Since the off-diagonal correlations decay exponentially for $t_N \gg (\beta T)^{-2}$, by numerically integrating the re-scaled results, we obtain for $N \gg (\beta T)^{-2}$,
\begin{equation}
\kappa_{od}( \infty) \approx 2.6\,\eta^2 (\beta T)^{-2}.
\end{equation}
%
%
Integrating Eq.~(\ref{eq:c_diagonal}),
we obtain the contributions from the diagonal correlations
\begin{equation}
\kappa_d(N) \approx \frac{2\eta}{N}  \sum_{k\ge k_m} \frac{v_k^2}{2\Gamma_k}+   \frac{\eta^2 k_m t_N}{\pi}  
\end{equation}
%
leading to the Peierls-Boltzmann relation for all modes $k\gtrsim  k_m$, whereas for the ballistic modes with $k\lesssim k_m$  we have used $e^{-2 \Gamma_k t_n }\approx 1-2 \Gamma_k t_N$. By using the expression Eq.~(\ref{eq:Gamma_k1}), we can obtain an explicit asymptotic expression
(see the supplementary material for details)
\begin{equation}    
\kappa_d(N) \propto \frac{5\, \left(\beta T \right)^{-4/5} \eta ^ 2}{2\pi\,(2 \Gamma_0)^{3/5}}   N^{2/5} + \mathcal{O}(N^{1/5}).
\label{kappa-asymp}
\end{equation}
%
%
%
Again, similar to the asymptotic expression of the autocorrelation function, leading and sub-leading orders are not easily distinguished when analyzing numerical data of finite system sizes. 
Let us further note that $2/5$ of the asymptotic value in Eq.~(\ref{kappa-asymp}) stems from contributions of the ballistic modes, $k\lesssim k_m$.
Therefore, although the power-law behavior is universal, the prefactor of $N^{2/5}$ depends on boundary conditions, so that noticeable deviations of the prefactor are expected in non-equilibrium simulations \cite{wang2011power}.
%
%
\paragraph{Discussion.} Our results provide a microscopic interpretation of the apparent change in transport behavior reported at weak anharmonicity.
As $\beta T$ decreases, the off-diagonal mode correlations persist up to a characteristic time, $\tau_{\mathrm{od}}\sim(\beta T)^{-2}$, delaying the crossover to the regime in which independent-mode contributions dominate the heat-current correlation. 
Consequently, progressively larger systems are required to resolve the long-time tail governed by the long-wavelength phonons, as precisely quantified here.
Beyond this crossover, we recover the same phonon-based reconstruction of the heat-current correlation and the same long-wavelength damping law, $\Gamma_k\sim k^{5/3}$, for all values of $\beta T$ considered. 

In conclusion, weak anharmonicity does {\em not} introduce a distinct asymptotic transport mechanism; instead, it merely shifts the onset of the phonon-dominated regime to longer times and larger length scales.
Our analysis also resolves the asymptotic scaling of the heat-current autocorrelation function, $C_J(t)\sim t^{-3/5}+ {\cal O}(t^{-4/5})$, which implies a conductivity divergence $\kappa(N)\sim N^{2/5} + {\cal O}(N^{1/5})$. This behavior is explained by the observed long-wavelength phonon damping $\Gamma_k\sim k^{5/3}(\beta T)^{4/3}$, although a rigorous analytical derivation of this scaling remains an open problem.
%
%
\bibliography{apssamp}
%
\end{document}


\title{Supplemental material}
\author{Kiratholly Nandakumar Madhav Sharma}
\affiliation{Univ. Grenoble Alpes, CNRS, LPMMC, 38000 Grenoble, France}
\author{Stefano Mossa}
\affiliation{Universit\'e Grenoble Alpes, CEA, IRIG-MEM-LSim, 38054 Grenoble, France}
\author{Jean-Louis Barrat}
\affiliation{Univ. Grenoble Alpes, CNRS, LIPhy, 38000 Grenoble, France}%
\author{Markus Holzmann}
\affiliation{Univ. Grenoble Alpes, CNRS, LPMMC, 38000 Grenoble, France}
\date{\today}
\maketitle

\section{Phonon Frequencies and Damping Rates}
\begin{figure}[h!]
\centering

\begin{subfigure}[t]{\columnwidth}
    \centering
    \includegraphics[
        width=\columnwidth,
        height=0.75\columnwidth
    ]{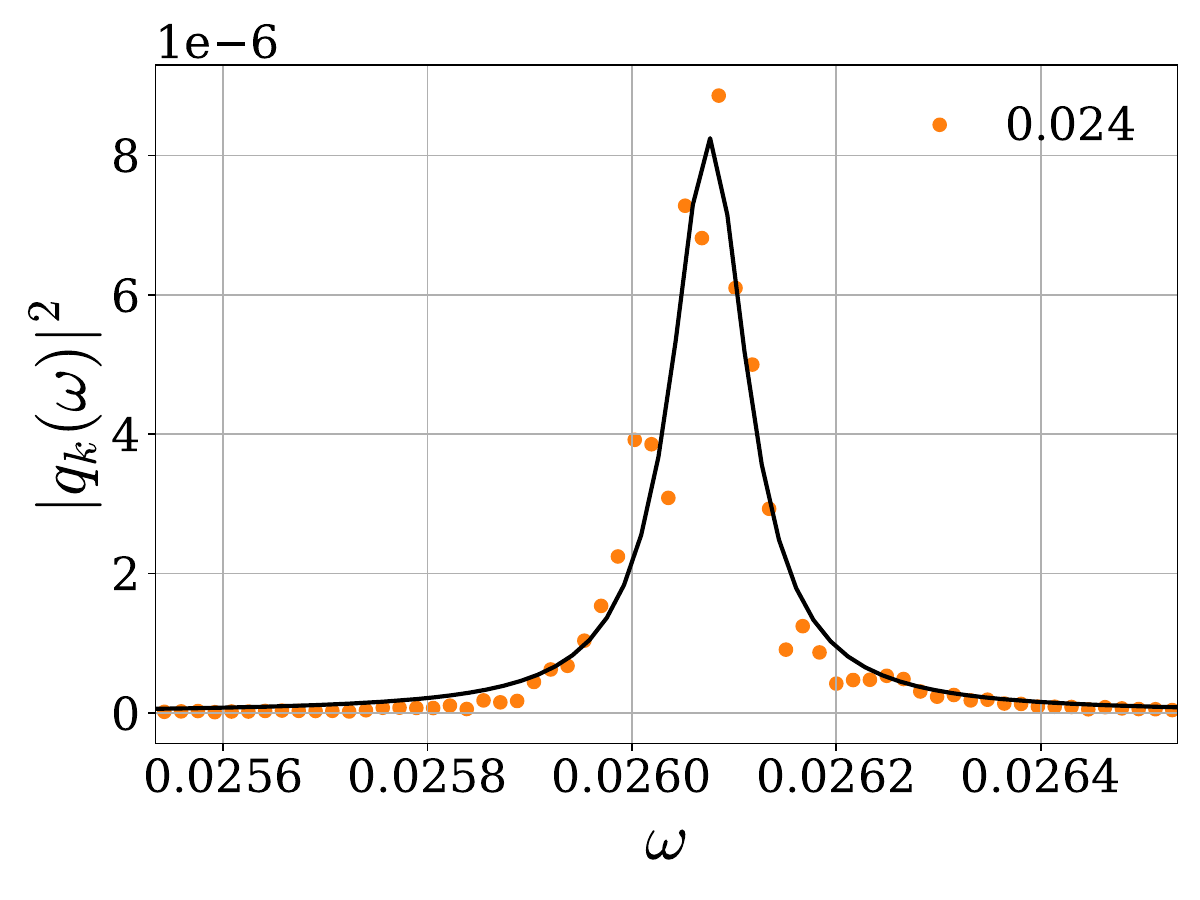}
    \captionsetup{justification=centering,singlelinecheck=true}
    \caption{}
    \label{fig:phonon_spectrum_betaT_0013}
\end{subfigure}
\vfill
\begin{subfigure}[t]{\columnwidth}
    \centering
    \includegraphics[
        width=\columnwidth,
        height=0.75\columnwidth
    ]{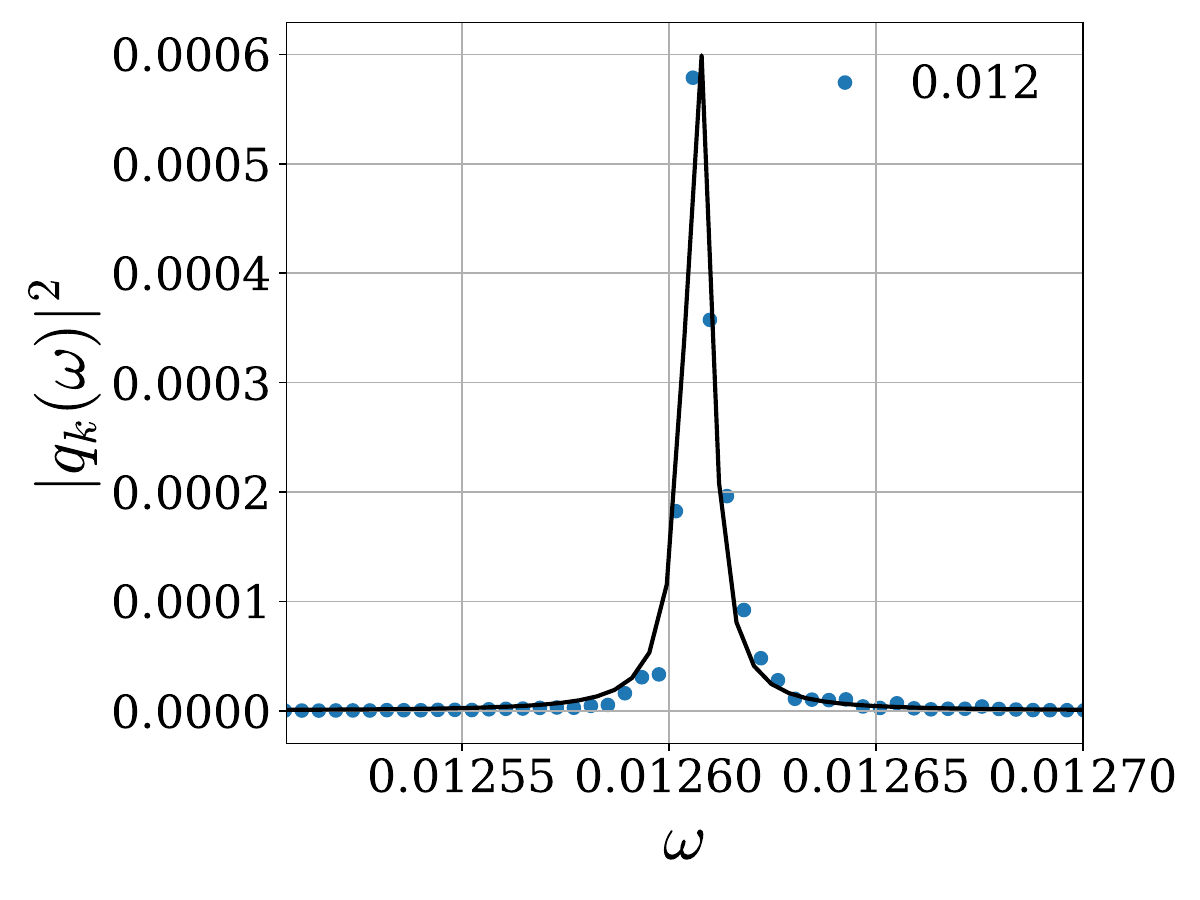}
    \captionsetup{justification=centering,singlelinecheck=true}
    \caption{}
    \label{fig:phonon_spectrum_betaT_0005}
\end{subfigure}

\caption{
    Representative phonon spectra with Lorentzian fits for $N=131072$ at different values of $\beta T$ and wave number $k$: (\subref{fig:phonon_spectrum_betaT_0013}) $\beta T=0.05$ and $k=0.024$ and (\subref{fig:phonon_spectrum_betaT_0005}) $\beta T=0.02$ and $k=0.012$. 
    Symbols denote the simulation data, and the solid line represents the Lorentzian fit.  
}
\label{fig:phonon_spectrum_examples}
\end{figure}

The phonon frequencies and lifetimes are extracted by fitting the phonon power spectrum to a Lorentzian line-shape function
\begin{equation}
|q_k(\omega)|^2
=
\frac{A_k}{(\omega-\omega_k)^2+\Gamma_k^2},
\label{eq:lorentzian_phonon_spectrum}
\end{equation}
where $\omega_k$ is the phonon frequency, $\Gamma_k$ is the spectral linewidth.
Fig.~\ref{fig:phonon_spectrum_examples} shows representative spectra from individual simulation runs, together with the corresponding Lorentzian fit. 

\begin{figure}[h]
    \centering
    \includegraphics[width=\columnwidth, height=0.85\columnwidth]{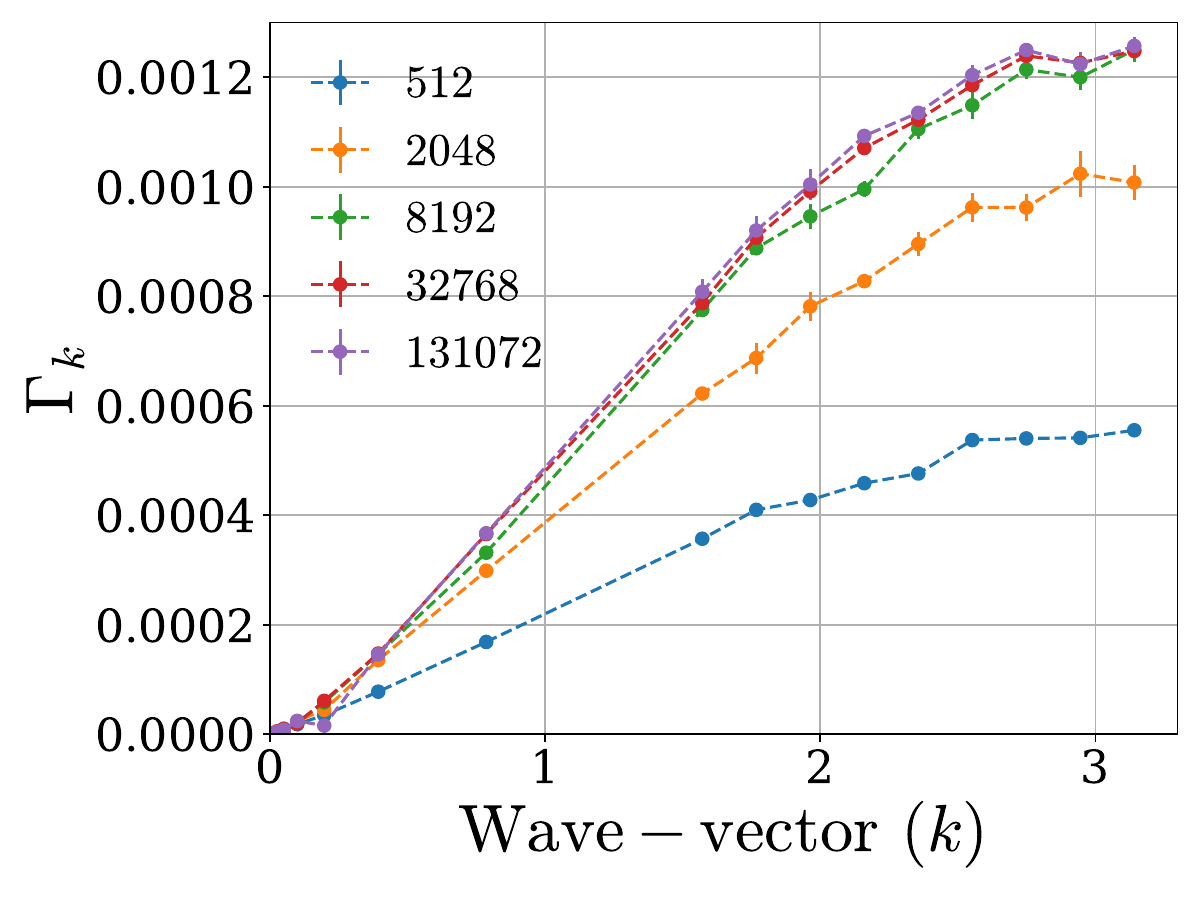}
    \caption{Phonon damping rates, $\Gamma_k$, as a function of the  wave-vector $k$ at the indicated values of  $N$ for $\beta T = 0.01$.}
    \label{fig:Phonon_lifetime_Vs_Wavevector_plot}
\end{figure}

\begin{figure}
    \centering
    \includegraphics[width=\columnwidth, height=0.85\columnwidth]{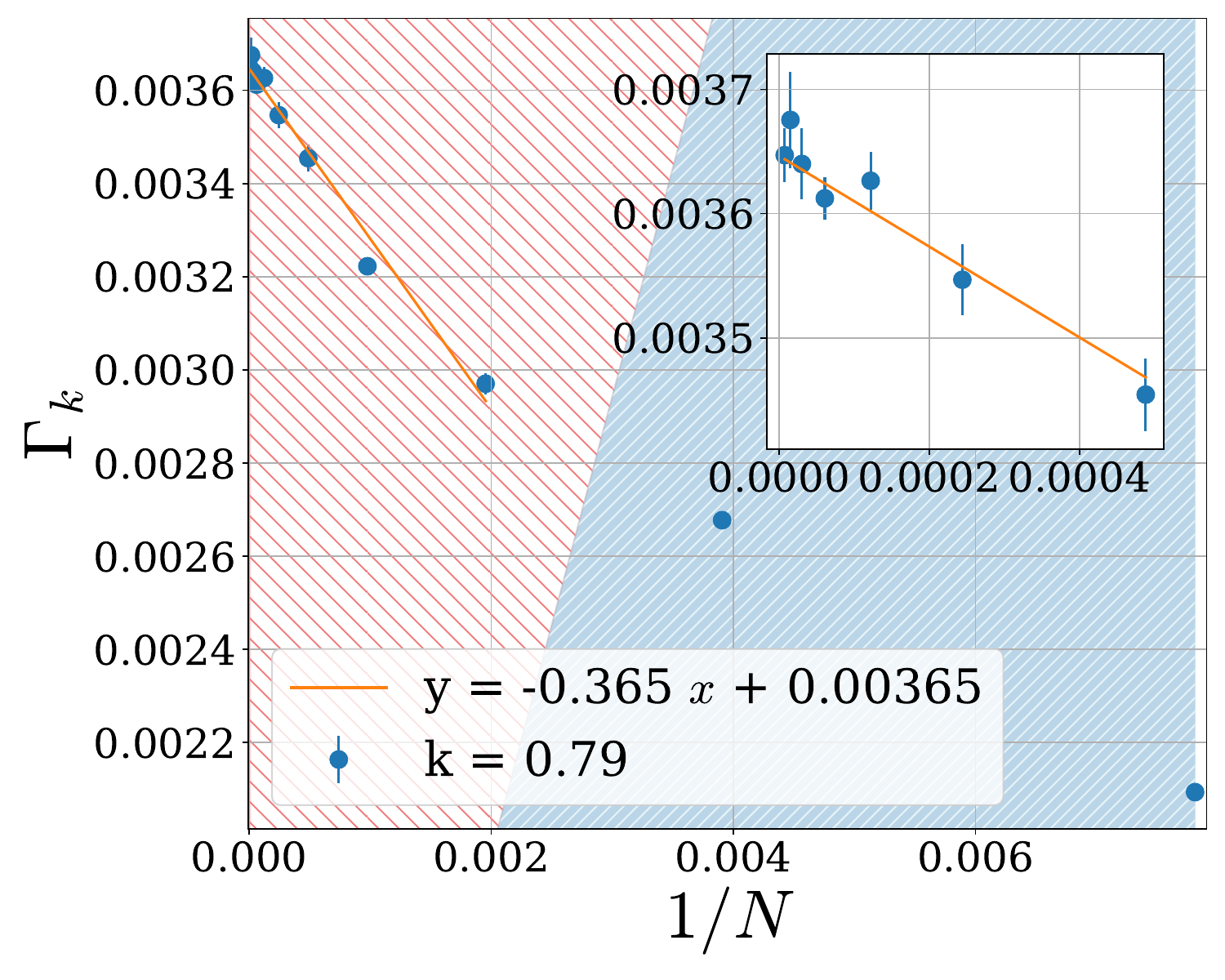}
    \caption{Phonon damping rate $\Gamma_k$ as a function of inverse system size $N^{-1}$ for $\beta T=0.10$. The wave vector $k$ and fitted parameter values are indicated in the plot. Error bars represent the standard error of the mean damping rate across independent simulation runs. The blue and red shaded regions correspond to phonon mean free path, $\ell_k>N$ and $\ell_k\leq N$, respectively. }
    \label{fig:Phonon_Damping_Rate_Vs_Inverse_N}
\end{figure}

We observe a strong finite-size effect on the phonon damping rate, as shown in Fig.~\ref{fig:Phonon_lifetime_Vs_Wavevector_plot}.
Each point represents the mean damping rate obtained by fitting the spectra of 8--10 independent runs separately. 
The error bars indicate the standard error of the mean. 
As illustrated in Fig.~\ref{fig:Phonon_Damping_Rate_Vs_Inverse_N}, we extrapolated the phonon damping rate for each mode to the thermodynamic limit ($N \rightarrow \infty$) assuming a linear dependence on $N^{-1}$ by restricting to system sizes  that exceed the phonon mean free path,
\begin{equation}
    N \gtrsim \ell_k = \frac{v_k}{\Gamma_k}.
\end{equation}

\section{Phonon Frequencies}
To leading order, the extracted phonon frequencies $\omega_k$ are proportional to the bare frequencies $\omega_k^0$, or \mbox{$\omega_k^2 \simeq \eta [\omega_k^0]^2$}.
From mean-field considerations \cite{PhysRevE.75.046603},
we expect a mode-independent proportionality constant $\eta$ approximately given by 
\begin{align}
    \eta = \frac{1 + \sqrt{1 + 12 \beta T}}{2},
    \label{eq:Renormalization_Constant}
\end{align}
\begin{figure}[b]
    \centering
    \includegraphics[width=\columnwidth, height=0.85\columnwidth]{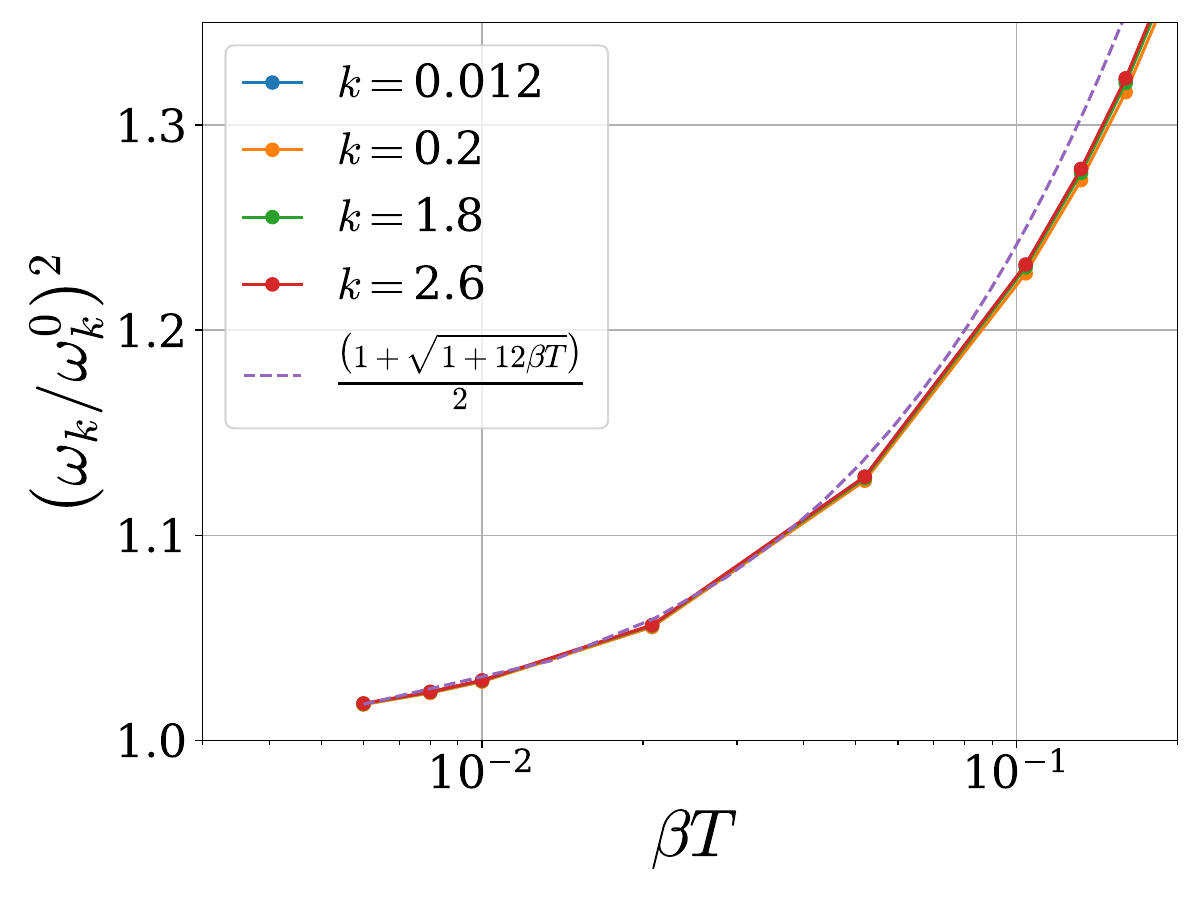}
    \caption{Ratio $(\omega_k/\omega_k^0)^2$, with $\omega_k$ the determined phonon frequencies and $\omega_k^0$ the bare harmonic frequencies, as a function of $\beta T$.
    We show data for $N=4096$ at the indicated values of the wave vector $k$. 
    The dashed line represents the approximate expression, Eq.~(\ref{eq:Renormalization_Constant}). }
    \label{fig:PhononFrequency}
\end{figure}
We can estimate the value of $\eta$ numerically by calculating the ratio $ (\omega_k/\omega^0_k)^2$ for any phonon mode, as shown in Fig.~\ref{fig:PhononFrequency}, where we show it for four different $k$ values.

\section{Total Heat-Current vs Harmonic Heat-Current}

The autocorrelation function of the total heat current, $J$,  differs from that of the bare harmonic heat current by a multiplicative factor approximately equal to $\eta^2$. 
This factor can be understood within an effective harmonic description, which can capture equilibrium observables even at moderate anharmonicity. 
In this description, the heat current retains its harmonic form, with bare phonon frequencies and group velocities replaced by their renormalized counterparts. 
Since both acquire a factor $\sqrt{\eta}$, the effective current acquires a multiplicative factor $\eta$ and its correlation function a factor $\eta^2$. 

\begin{figure}[t]
    \centering
    \includegraphics[width=\columnwidth, height=0.9\columnwidth]{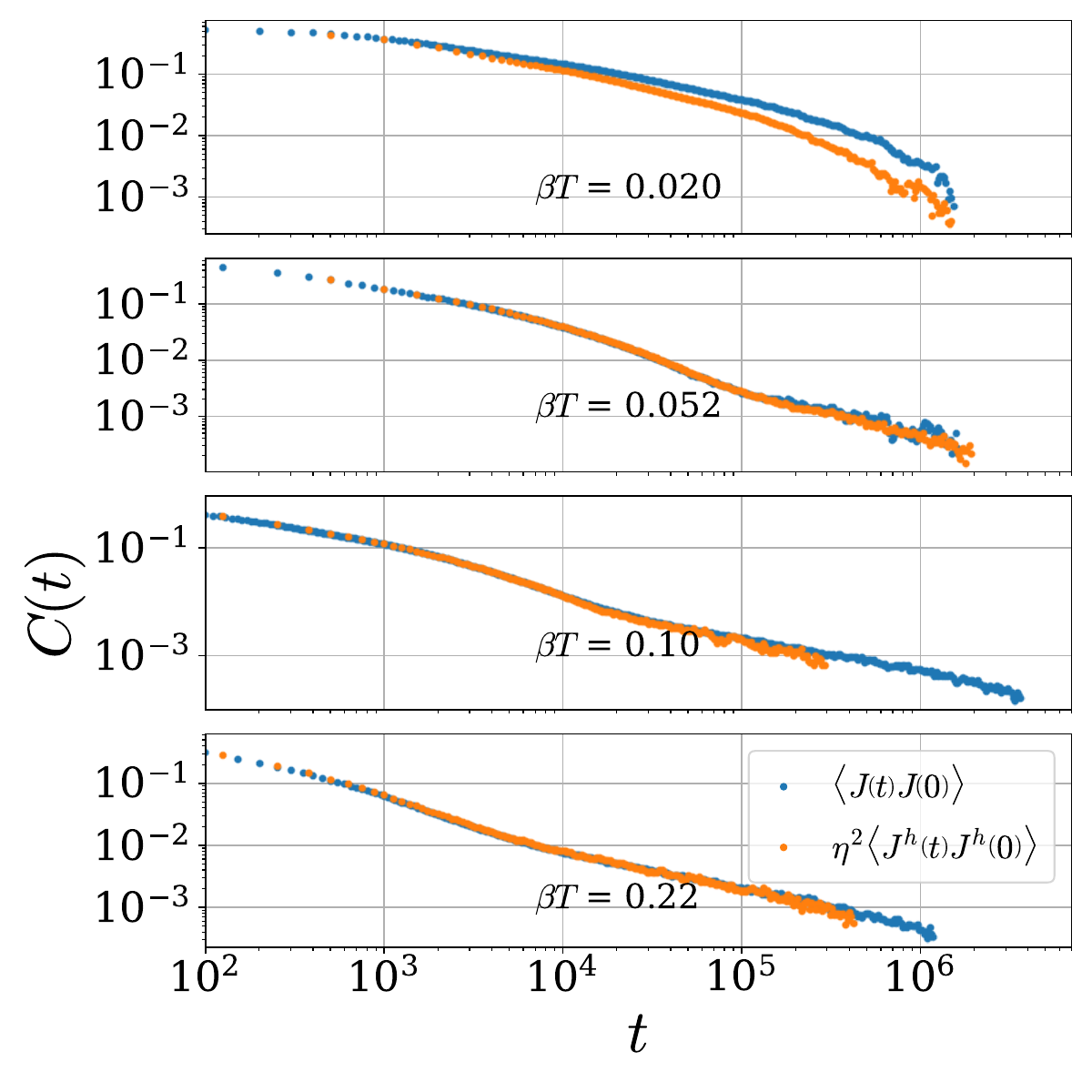}
    \caption{
    Comparison between the total heat-current autocorrelation function, $\langle J(t)J(0) \ \rangle$, and the rescaled harmonic contribution, $\eta^2 \langle J^h(t)J^h(0)\rangle$, orange and blue symbols, respectively, for \(N=8192\) and four different $\beta T$ values as specified in the plots.
    The data are logarithmically binned using $100$ bins per decade for visualization. 
    Each marker represents the geometric mean of the times and the median correlation value within its bin.
    }
    \label{fig:total_vs_harmonic_hcacf}
\end{figure}

In Fig.~\ref{fig:total_vs_harmonic_hcacf}, we  compare our data of the total current correlation function with the autocorrelation function of the effective harmonic currents, $\eta^2 \langle J^h(t) J^h(0)\rangle$.
This comparison shows four values of $\beta T$, and the same relation holds across all considered anharmonicity strengths.

\section{Relation Between Phonon and Mode-Current Damping Rates}
\begin{figure}[t!]
    \centering
    \includegraphics[width=\columnwidth]{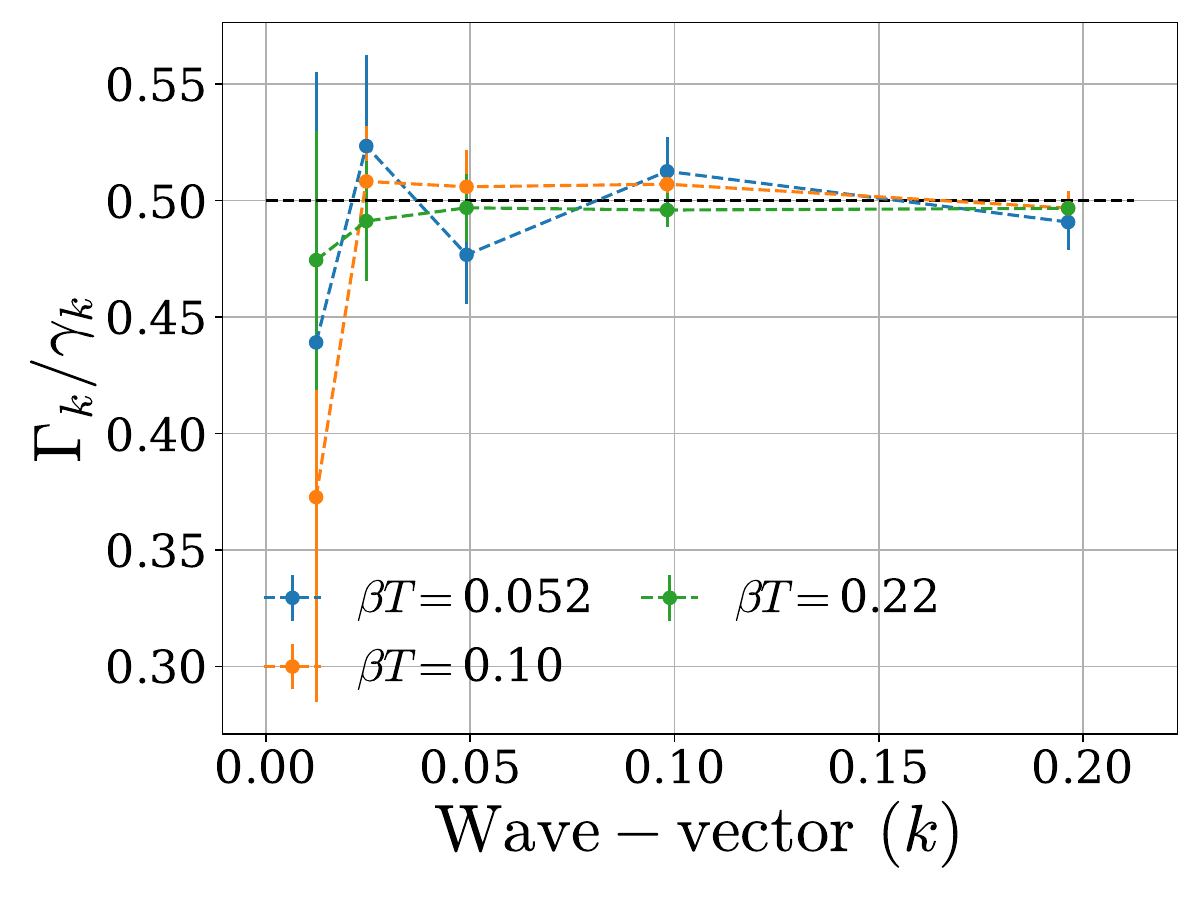}
    \caption{Ratio of the phonon damping rate $\Gamma_k$ to the mode-current damping rate $\gamma_k$ as a function of the wave number $k$, for $N=8192$ and three values of $\beta T$.
    The horizontal dashed line denotes the theoretical prediction $\Gamma_k/\gamma_k=1/2$.}
    \label{fig:Ratio_of_Phonon_Damping_Rate_over_Mode_current}
\end{figure}

In Figure 5 of the main text, we show that the modes of the current-current correlations individually decay exponentially,  $C_k(t) \sim \langle J_k(t) J_k(0)\rangle \sim e^{- \gamma_k t}$, and collapse to a single
function of $\Gamma_k t$, where $\Gamma_k$ are given
by the phonon damping rates, leading to a damping rate of
$\gamma_k=2 \Gamma_k$.
In 
Fig.~\ref{fig:Ratio_of_Phonon_Damping_Rate_over_Mode_current} we directly plot the ratio of the individually extracted damping rates, $\gamma_k$, with the
phonon damping rates, $\Gamma_k$ at a fixed system size
$N = 8192$ for different
anharmonicities $\beta T$ and wave vectors.   
The obtained ratios are consistent 
with $\gamma_k=2\Gamma_k$.

\section{Heat Current Expression in the Thermodynamic Limit}
We can obtain the expression for $C_{d}^{\infty} (t)$ given in the main text starting from 
\begin{equation}
    C_d(t) \simeq \frac{2\eta}{N} \sum_{k > 0}^{\pi} v_k^{2}e^{-2 \Gamma_k t}.
\end{equation}
In the thermodynamic limit, we can replace the summation with an integral, 
\begin{equation}
     C_d^{\infty} \left(t\right) = 2 \eta \int_0^\pi \frac{dk}{2\pi} \, v_k^2 e^{-2 \Gamma_k T}.
     \label{eq:Heat_Current_Correlation_in_Thermodynamic_limit}
\end{equation}
Using for $\Gamma_k$ now the approximate expression
\begin{equation}
    \Gamma_k \approx
\Gamma_0 (\beta T)^{4/3} k^{5/3}\left(1 - (k/k_0)^{1/3}\right) 
\label{eq:Binomial_expansion_of_Phonon_Damping_Model}
\end{equation}
as well as $v_k^2 = \eta^2 \cos \left(\frac{k}{2}\right)$,
 we get 
\begin{equation}
    C_{d}^{\infty}\left(\tau\right) = \frac{\eta^2}{\pi} \int_{0}^{\pi} \cos^2\left(\frac{k}{2}\right) e^{-k^{5/3}\tau} e^{k_0^{1/3} k^2\tau}.
\end{equation}
where we defined $\tau = t/t_0$ where $t_0 = \frac{1}{2 \Gamma_0 \left(\beta T\right)^{4/3}}$.

We now perform a change of variable $x = k\tau^{3/5}$, and approximate the term $e^{k_0^{1/3} x^2 \tau^{-1/5}}$ and $\cos^2\left(\frac{x}{2\tau^{3/5}}\right)$
in the limit $\tau \rightarrow \infty$ (or $t \gg t_0$)  by their respective Taylor series expansions. Keeping only the leading-order terms, we get 
\begin{equation}
    C_{d}^{\infty}\left(\tau\right) \simeq \frac{\eta^2}{\pi} \int_{0}^{\infty}  e^{-x^{5/3}} \tau^{-3/5} \, dx + \mathcal{O}\left(\tau^{-4/5} \right)
    \label{eq:Heat_Current_Correlation_in_Thermodynamic_limit_with_change_of_variable}
\end{equation}
where we have put the upper bound of the integral to infinity.
The integral in Eq.~\ref{eq:Heat_Current_Correlation_in_Thermodynamic_limit_with_change_of_variable} can then be solved using  
\begin{equation}
    \int_{0}^{\infty} x^n e^{-x^{5/3}}\, dx =  \frac{3}{5}\Gamma\left(\frac{3(n + 1)}{5}\right).
\end{equation}
and we finally obtain the expression given in the main text 
\begin{equation}
   C_{d}^{\infty}\left(t \gg t_o\right) = \frac{3\eta^2}{5 \pi} \Gamma\left(\frac{3}{5}\right)\left(\frac{t_0}{t}\right)^{3/5}
   + \, \mathcal{O}\left(\left(\frac{t_0}{t}\right)^{4/5} \right) 
\end{equation}

\section{Off-Diagonal Contribution to Thermal Conductivity}
To quantify the contribution of cross-mode correlations to thermal transport, we decompose the thermal conductivity as
\begin{align}
    \kappa = \kappa_{d} + \kappa_{od},
\end{align}
where the off-diagonal contribution is given by
\begin{align}
    \kappa_{od}(\infty) = \lim_{\tau \rightarrow \infty}\int_{0}^{\tau} C_{od}(t)\,dt.
\end{align}
where 
\begin{equation}
    C_{od}(t) =\frac{1}{NT^2}\sum_{k \neq k^{\prime}} \left\langle J_k(t)J_{k^{\prime}}(0)\right\rangle.
\end{equation}

\begin{figure}[h]
    \centering
    \includegraphics[width=\columnwidth, height=0.85\columnwidth]{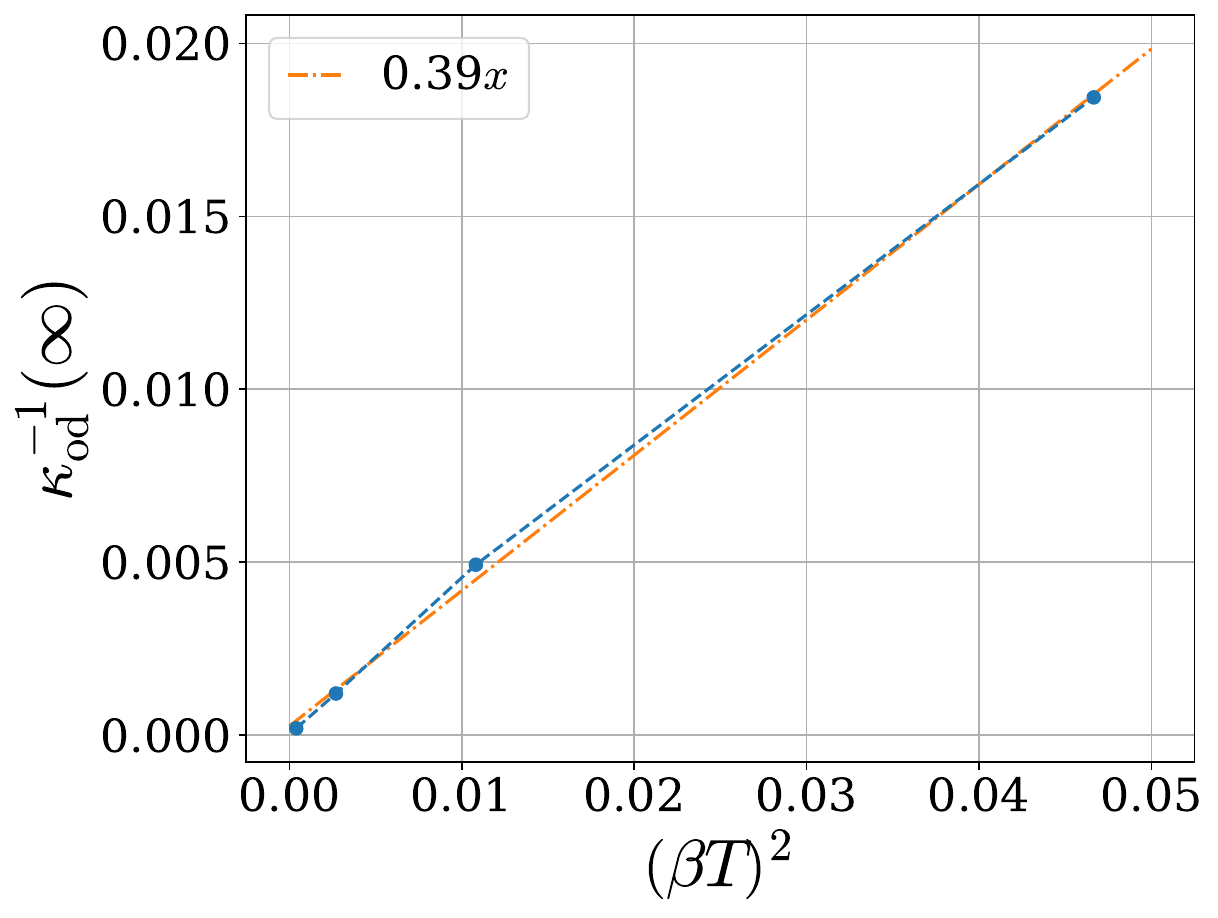}
    \caption{Inverse off-diagonal thermal conductivity, $\kappa_{\rm od}^{-1}(\infty)$, as a function of $(\beta T)^2$.
    Data are obtained from $N=8192$ simulations for $\beta T=0.216$, $0.104$, $0.052$ and $0.020$.
    }
    \label{fig:off_diagonal_contribution_scaling_with_respect_to_betaT}
\end{figure}

We plot $\kappa_{od}^{-1}(\infty)$ as a function of $(\beta T)^2$ in Fig.~\ref{fig:off_diagonal_contribution_scaling_with_respect_to_betaT}. 
Note that, by construction, $C_{od}$ starts from zero and rises to a maximum value before exhibiting exponential decay, and the integration grid to calculate $\kappa_{od}(\infty)$ should be small enough to resolve this peak.  
The inverse of $\kappa_{od}(\infty)$ exhibit an approximately quadratic dependence in $(\beta T)^2$, with a prefactor of $0.39$ independent of system size, 
see  Fig.~\ref{fig:off_diagonal_contribution_scaling_with_respect_to_betaT},
of $\kappa_{od} (\infty) \approx 2.56(\beta T)^{-2}$. 

\section{Thermal Conductivity Calculation}
We have shown that $\kappa$ for a given $\beta$-FPUT lattice can be expressed as the sum of diagonal $(\kappa_d)$ and off-diagonal $(\kappa_{od})$ contributions from the harmonic heat-current autocorrelation function.
Whereas the off-diagonal contributions reach a finite asymptotic value,  the diagonal contributions diverge as the system size increases, approximately given by
\begin{align}
    \kappa_{d}(N) \approx& \frac{2\eta}{N}  \sum_{k\ge k_m} \frac{v_k^2}{2\Gamma_k}+   \frac{\eta^2 k_m t_N}{\pi}  
\end{align}
where $k_m \simeq (2\Gamma_0)^{-3/5}\left(\beta T\right)^{-4/5}N^{-3/5}$, 
\begin{align}
    \kappa_{d}  =& \frac{\eta^2 \left(\beta T \right)^{-4/3}}{2\pi\Gamma_0}  \int_{k_m}^{\pi} dk \, \frac{\cos^2(k/2)}{k^{5/3}} \nonumber \\
    &+ \frac{\eta^2 \left(\beta T\right)^{-4/5}}{\pi \left(2 \Gamma_0\right)^{3/5}} N ^{2/5} + \mathcal{O} \left(N^{1/5}\right)
    \label{eq:kappa_d_expression_before_integral_approximation}
\end{align}
Within leading order, the integral is given by
\begin{eqnarray}
    \int_{k_m}^{\pi} dk\,\frac{\cos^2(k/2)}{k^{5/3}}
    &\approx&
    \frac{3}{2}(\beta T)^{8/15}(2\Gamma_0)^{2/5} N^{2/5}
    \nonumber\\
    && {} + \mathcal{O}\!\left(N^{1/5}\right).
    \label{eq:Integral_approximation}
\end{eqnarray}
so that we get 
\begin{align}
    \kappa_d = \frac{5\eta^2 \left(\beta T \right)^{-4/5}}{2\pi(2 \Gamma_0)^{3/5}} N^{2/5} + \mathcal{O} \left(N^{1/5}\right).
\end{align}

\bibliography{apssamp}